\documentclass[aps,pra,notitlepage,twocolumn,10pt,a4paper]{revtex4-1}

\usepackage{xcolor,graphicx,ulem}
\usepackage{amsmath,amssymb}
\usepackage[colorlinks=true,urlcolor=blue,citecolor=blue,linkcolor=blue]{hyperref}

\begin{document} 

\title{Preparation of arbitrary quantum states with regular $P$~functions}

\author{B. K\"uhn}\email{benjamin.kuehn2@uni-rostock.de}
\author{W. Vogel}
\affiliation{Arbeitsgruppe Quantenoptik, Institut f\"ur Physik, Universit\"at Rostock, D-18051 Rostock, Germany}

\date{\today}

\begin{abstract} 
	We propose a quantum optical device to experimentally realize quantum processes, which perform the regularization of the---in general highly singular---Glauber--Sudarshan $P$~functions of arbitrary quantum states before their application and/or measurement.
	This allows us to produce a broad class of nonclassical states with regular $P$~functions, also called nonclassicality quasiprobabilities.
	For this purpose, the input states are combined on highly transmissive beam splitters with specific Gaussian or non-Gaussian classical states.
	We study both balanced and unbalanced homodyne detections for the direct sampling of the output states of the implemented processes, which requires no further regularization or state-reconstruction.
	By numerical simulations we demonstrate the feasibility of our approach and we outline the generalization to multimode light.
\end{abstract}

\maketitle

\section{Introduction}
\label{ch:introduction}

	Quantum processes, that convert light from one state to another, are fundamental building blocks in modern quantum information processing \cite{Nielsen2000}.
	In particular, the implementation of specific processes in optical devices to produce states with desired properties out of available resources has become an increasingly important subject in quantum optics.
	One can distinguish classical from nonclassical processes on the basis of Titulaer and Glaubers definition of nonclassicality \cite{Titulaer1965}, which is based on the Glauber-Sudarshan $P$~function \cite{Glauber1963,Sudarshan1963}.
	The former transfer classical states to classical ones, while the latter can map a classical input to a nonclassical output; see also Ref. \cite{Rahimi2013}.
	Prominent examples for nonclassical processes, already realized by established techniques, are, e.g., photon addition \cite{Zavatta2007}, squeezing \cite{Slusher1985,Wu1986}, and entanglement generation \cite{Einstein1935,Schroedinger1935,Roslund2014}.
	Nevertheless, classical processes such as coherent displacement~\cite{Wallentowitz1996}, optical phase randomization \cite{Franzen2006,Kiesel2009}, and photon-subtraction \cite{Ra2017,Averchenko2016} also find numerous technical applications. 
	Interestingly, photon subtraction can even enhance the nonclassical effect of quantum entanglement \cite{Horodecki2009,Walschaers2017}.
	
	An important tool for the appropriate characterization and visualization of quantum states, produced by a given process, are phase-space functions \cite{Agarwal1970,Cahill1969,Wigner1932}, also known as quasiprobabilities. 
	The $P$~function is the key quasiprobability for the certification of nonclassical effects, since its negativities unambiguously uncover the nonclassicality of quantum states.
	However, this function is in general highly singular and not experimentally accessible.
	Accordingly, it is often hard to infer from measurements that it fails to have the properties of a classical probability distribution.
	Other quasiprobabilities are smoothed versions of the $P$~function and are regular functions for sufficiently strong smoothing.
		
	Specific quasiprobabilities, the so-called nonclassicality quasiprobabilities \cite{Kiesel2010}, were designed to uncover all possible nonclassical effects through their negativities.
	These always regular functions provide a universal nonclassicality test. 
	The long-known $s$-parametrized quasiprobabilities \cite{Cahill1969}, on the other hand, only certify nonclassical properties of a subclass of states.
	Note that the $s$-parametrized quasiprobabilities cannot visualize the nonclassicality of squeezed states because they are either singular or non-negative.
	The technique of nonclassicality quasiprobabilities has been successfully applied in several experiments; see Refs. \cite{Agudelo2015,Kiesel2012_5,Kiesel2011_2,Kiesel2011_1}.
	Moreover, it has also been extended to multimode light \cite{Agudelo2013}.
	Quite recently it was shown that the nonclassicality quasiprobabilities even yield detailed insight into the structural characteristics of quantum states.
	In particular, they directly include information on both quantum non-Gaussianity and the degree of nonclassicality~\cite{Kuehn2018}, whose study so far required additional techniques, such as those in Refs.~\cite{Jezek2011,Gehrke2012,Sperling2014}, respectively.
	
	In the previous contributions, the regularization of the $P$~function was implemented after the measurement as a numerical filtering of the recorded data, while the $P$~function of the state under study may be a singular one. 
	Hence, the sampled nonclassicality quasiprobability does not represent the quantum state of the probed light in a coherent-state basis.
	This raises the following question: Is it possible to perform the regularization procedure also before any measurement, such that the $P$~function of the resulting state is in principle experimentally accessible?
	To our knowledge, it is an open question yet, whether there exists a single optical device, realizing a specific quantum process, that fulfills the following requirements: 
	First, it converts an arbitrary input quantum state with a possibly singular $P$~function to an output state with a regular $P$~function (condition I).
	Second, all nonclassical (classical) input states are mapped to nonclassical (classical) output states (condition II).
	Such an apparatus provides us with a broad class of nonclassical states with a regular $P$~function, hence we overcome the fundamental difficulty that the $P$~function is in general not experimentally accessible.
	Beneficially, the nonclassicality can in this case be directly demonstrated---on the basis of the original definition---by negative values of the measurable $P$~function of the generated state, without the need to apply nonclassicality criteria, which are only sufficient but not necessary.
	Furthermore, the generated light is then available for quantum applications or as a resource for further state manipulations through other quantum processes.
	
	So far only few examples of nonclassical states with regular $P$~function are known.
	Prominent examples are the photon-added thermal states \cite{Agarwal1992}, which are obtained by exciting a classical thermal field by a certain number of photons \cite{Zavatta2007}.
        More recently, another technique was proposed to produce such kinds of states by puncturing classical states \cite{Damanet2017}.
        In both cases condition II is violated, since classical states are mapped to nonclassical states.
        Exposing a state to sufficiently strong thermal noise can regularize the $P$ function. 
	The resulting states are due to the controllable degree of thermal noise, which washes out nonclassical effects, useful to test the power of different nonclassicality criteria. 
	Unfortunately, this technique does not yield the desired outcome (conditions~I and~II) for all input states.
	It is an interesting problem to find alternative ways to perform the regularization of the $P$~function in experiments without converting nonclassical to classical states or vise versa, such that both conditions~I and~II are satisfied for all possible input states.
		
	In the present paper, we show that the mapping of states with a singular $P$~function to another one with its $P$~function representing a well-behaved nonclassicality quasiprobability can be implemented in experiments while preserving the nonclassicality (classicality).
	This requires linear optical elements as well as non-Gaussian classical light sources.
	The feasibility of this technique is demonstrated for a single-mode squeezed vacuum state via numerical simulations and we verify the nonclassicality on the basis of the $P$~function sampled from the data of a simulated balanced homodyne detection.
	Our method works for both discrete and continuous variable regimes and is straightforwardly generalized to multimode scenarios.
	
 	Our paper is structured as follows.
 	In Sec. \ref{ch:regularization}, we define the term nonclassicality used throughout this work and recall quasiprobabilities and associated numerical regularization strategies for the Glauber-Sudarshan $P$~function. 
 	Furthermore, we explain how these quantities are accessible via both balanced and unbalanced homodyne detection.
 	An experimental scheme is proposed in Sec. \ref{ch:experiment}, which enables a quantum optical implementation of the filtering process.
 	We also extend the method to multimode systems.
 	The verification of the resulting regularized states by sampling their $P$~functions is considered in Sec. \ref{ch:measurement} and demonstrated for the example of a squeezed vacuum state via simulated data.
 	We summarize and conclude in Sec. \ref{ch:conclusions}.
	
\section{Nonclassicality in terms of $P$~functions} 
\label{ch:regularization}

	Among the abundance of pure quantum states, the coherent states $|\alpha\rangle$ most resemble the classical properties of light.
	Mixing two or more of these coherent states with different amplitudes $\alpha\in\mathbb{C}$ is merely a statistical distribution of different harmonic-oscillator amplitudes, which can be properly described in the framework of classical statistical physics.
	It is, therefore, reasonable to define a classical state $\hat\rho_{\mathrm{cl}}$ as one which can be written as a mixture of coherent states,
	\begin{align}\label{eq:clstate}
		\hat\rho_{\mathrm{cl}}=\int d^2\alpha\,P_{\mathrm{cl}}(\alpha)\,|\alpha\rangle\langle\alpha|,
	\end{align}	
	with a classical probability distribution $P_{\mathrm{cl}}$.
	More generally, any quantum state can be represented as \cite{Glauber1963,Sudarshan1963}
	\begin{align}\label{eq:GS}
		\hat\rho=\int d^2\alpha\,P(\alpha)\,|\alpha\rangle\langle\alpha|,
	\end{align}
	with the Glauber--Sudarshan $P$~function.
	Since this phase-space function can violate the properties of a classical probability distribution, it provides---compared with Eq. \eqref{eq:clstate}---the basis for a definition of nonclassicality.
	In particular, if there is no $P$~function representing a given state $\hat\rho$ with the properties of a classical probability distribution, and, thus, this state cannot be written as in Eq. \eqref{eq:clstate}, it is referred to as nonclassical \cite{Titulaer1965}.

\subsection{Regularized $P$~functions}	
	A central problem associated with the function $P(\alpha)$ is its high singularity for a huge class of quantum states.
	For example, the $P$~function of the squeezed vacuum state, 
	\begin{align}\label{eq:sqstate}
		|\xi\rangle=\exp\left[\tfrac{1}{2}\left(\xi^\ast\hat a^2-\xi\hat a^{\dagger 2}\right)\right]|0\rangle,
	\end{align}
	with the squeezing parameter $\xi\in\mathbb{C}\setminus \{0\}$ and the bosonic annihilation (creation) operator $\hat a$ ($\hat a^\dagger$), contains infinite derivatives of the Dirac delta distribution \cite{Kiesel2011_2,Sperling2016}.
	It is, therefore, not possible to reconstruct this function directly out of experimental data.
	In the following we use the short-hand notion ``regular states'' (``irregular states'') for states with regular (singular) $P$~functions.
	
	To resolve this deficiency, one considers smoothed versions of the $P$~function, still completely characterizing a quantum state.
	These quasiprobabilities can be regular functions, and are, thus, experimentally accessible.
	Gaussian filtering leads to the class of $s$-parametrized quasiprobabilities \cite{Cahill1969}, which, however, do not provide for all nonclassical states a regular phase-space function attaining negative values.
	
	To the best of our knowledge, the filtering method that allows us to uncover all nonclassical effects, is provided by so-called nonclassicality filters $\Omega_w(\beta)$; see Ref. \cite{Kiesel2010}.
	The latter map the $P$~function to a new phase-space function  
	\begin{align}\label{eq:convolution}
		P_w(\alpha)=\int d^2\gamma\,\tilde \Omega_w(\gamma)\,P(\alpha-\gamma),
	\end{align}
	conserving the normalization $\int d^2\alpha\,P_w(\alpha)=1$, through a convolution with the Fourier transform,
	\begin{align}\label{eq:ftfilter}
		\tilde\Omega_w(\gamma)&=\dfrac1{\pi^2}\int d^2\beta\,\Omega_w(\beta)\,e^{\gamma\beta^\ast-\gamma^\ast\beta},
	\end{align}
	of the filter.
	To avoid additional negativities in $P_w$ which are introduced by the filtering itself, the nonclassicality filters have to have a non-negative Fourier transform $\tilde\Omega_w$.
	The real positive parameter $w$ controls the degree of smoothing.
	In the limit $w\to\infty$ one obtains the original $P$~function.
	The nonclassicality filters are constructed in such a way that, for all finite $w$, the function $P_w$ is regular. 
	It was shown that the quasiprobabilities $P_w$ provide a complete nonclassicality test.
	In fact, a necessary and sufficient condition for a given state to be nonclassical is that $P_w(\alpha)<0$ for some $\alpha$ and $w$; see Ref. \cite{Kiesel2010}.
	
\begin{figure}[h]
		\includegraphics[clip,scale=0.9]{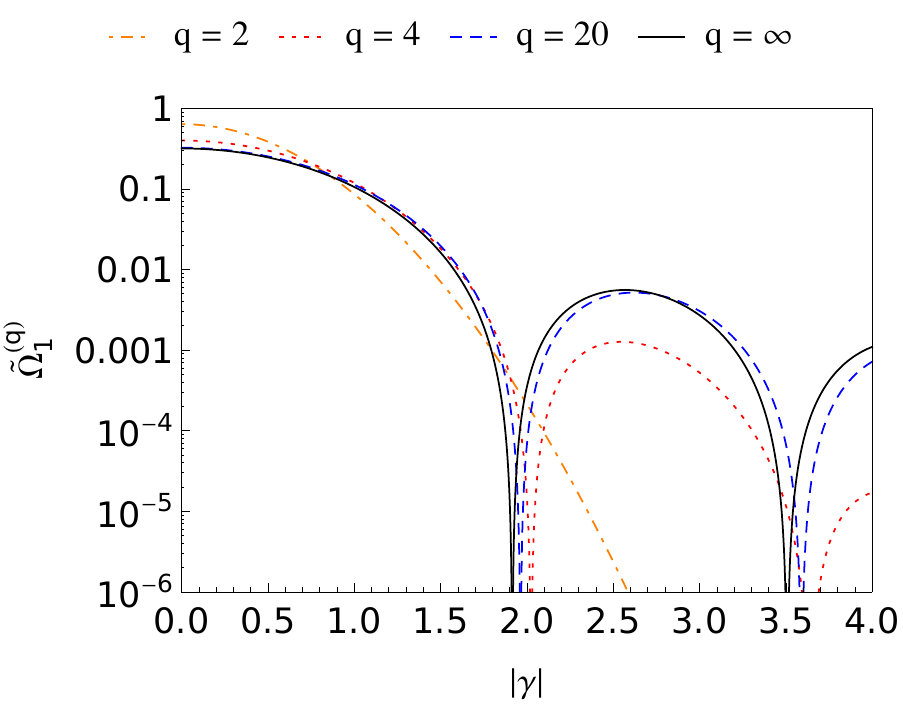}
		\caption{(Color online)
		Logarithmic plot of the Fourier transform $\tilde \Omega^{(q)}_w$ of the filters $\Omega^{(q)}_w$ as a function of $|\gamma|$ with filter parameter $w=1.0$ and for various values of the parameter $q$.
 	}\label{fig:ft}
	\end{figure}

	The specific filter functions, which fulfill these requirements are, e.g., the non-Gaussian autocorrelation filters \cite{Kiesel2010}
	\begin{align}\label{eq:Omegaq}
		\Omega^{(q)}_w(\beta)=\int d^2\beta'\,\omega^{(q)\ast}_w(\beta')\omega^{(q)}_w(\beta+\beta'),
	\end{align}
	with the infinitely differentiable function
	\begin{align}\label{eq:smallomega}
		\omega^{(q)}_w(\beta)=\dfrac1{w}2^{1/q}\sqrt{\dfrac{q}{2\pi\Gamma(2/q)}}\exp\left[-\left(\dfrac{|\beta|}{w}\right)^q\right]
	\end{align}
	and $2<q<\infty$.
	Here $\Gamma(\cdot)$ is the gamma function.
	Note that Gaussian filtering corresponds to the limiting case of $q=2$.
	For the opposite limiting case $q\to\infty$, the filter function is given by the analytical expression \cite{Kuehn2014}
	\begin{align}\label{eq:usedfilter}
		\Omega^{(\infty)}_w\!(\beta)\!=\!\dfrac{2}{\pi}\!\left[\mathrm{arccos}\left(\dfrac{|\beta|}{2w}\right)-\dfrac{|\beta|}{2w}\sqrt{1-\dfrac{|\beta|^2}{4w^2}}\,\right]\!\mathrm{rect}\left(\dfrac{|\beta|}{2w}\right)\!,
	\end{align}
	with $\mathrm{rect}(x)=1$ for $|x|\leq 1$, and zero otherwise.
	Figure~\ref{fig:ft} visualizes the Fourier transforms $\tilde\Omega_w^{(q)}$ of these filters for $w=1.0$ and for various values of the  parameter $q$.
	The functions for other parameters $w$ are easily obtained by the rescaling transform $\tilde\Omega^{(q)}_w(\gamma)=w^2\tilde\Omega^{(q)}_1(w\gamma)$.
	
	The usefulness of quasiprobabilities to certify nonclassicality has been demonstrated for various states \cite{Kiesel2010,Biswas2007}.
	Moreover, different experimental schemes for the reconstruction of quasiprobabilities have been analyzed \cite{Agudelo2015,Kuehn2016,Kiesel2012_2,Kiesel2012_3,Wallentowitz1996,Smithey1993,Vogel1989}.
	Among those, we consider the balanced and the unbalanced homodyne detection in the following.
	
\subsection{Balanced homodyne detection of irregular states}
	The balanced homodyne detection records the statistics $p(x;\phi)$ of the quadrature, $\hat x_\phi=\hat a e^{i\phi}+\hat a^{\dagger} e^{-i\phi}$, of a particular optical phase $\phi$; see also \cite{Grabow1993}.
	In Refs. \cite{Kiesel2011_2,Agudelo2015} it was shown that the quasiprobability $P_w$ is obtained out of $N$ quadrature-phase data points $\left\{(x_j,\phi_j)\right\}_{j=1}^N$, with phases $\phi_j$, uniformly distributed in the interval $[0,2\pi]$, by the sampling formula
	\begin{align}\label{eq:Pwsamplingbalanced}
		P_w(\alpha)\approx \dfrac1{N}\sum_{j=1}^N f_w(\Lambda_{j,\alpha}),
	\end{align}
	with the pattern function
	\begin{align}\label{eq:f}
		f_w(\Lambda_{j,\alpha})=\dfrac{2}{\pi}\int_0^\infty db\,b\,e^{b^2/2}\Omega_w(b)\,\cos\left(\Lambda_{j,\alpha}b\right),
	\end{align}
	and
	\begin{align}\label{eq:Lambda}
		\Lambda_{j,\alpha}=x_j+2|\alpha|\sin\left[\arg(\alpha)+\phi_j-\pi/2\right].
	\end{align}
	Here we assumed filter functions $\Omega_w(\beta)$ which only depend on the modulus $b=|\beta|$ rather then the argument of $\beta$.
	Furthermore, the product $\Omega_w(|\beta|) e^{|\beta|^2/2}$ has to be square integrable, which is fulfilled, e.g., for the autocorrelation function with $q>2$ in Eq. \eqref{eq:Omegaq} together with Eq. \eqref{eq:smallomega}.
	As the sampling points are measured independently, the associated statistical error of \eqref{eq:Pwsamplingbalanced} is given by
	\begin{align}\label{eq:sigma}
		\sigma\left\{P_w(\alpha)\right\}=\dfrac1{\sqrt{N(N-1)}}\sqrt{\sum_{j=1}^N\left[f_w(\Lambda_{j,\alpha})-P_w(\alpha)\right]^2}.
	\end{align}

\subsection{Unbalanced homodyne detection of irregular states}	
	Another possibility to determine quasiprobabilities is to measure the coherently displaced photon-number statistics, $p_n(\alpha)$, via unbalanced homodyne detection; see \cite{Wallentowitz1996}.
	It essentially combines the signal field with a coherent state of a particular amplitude on a highly transmissive beam splitter.
	This procedure coherently displaces the signal state $\hat \rho$ by a certain amplitude $\alpha$; that is, the resulting state is $\hat\rho(-\alpha)=\hat D^\dagger(\alpha)\hat\rho \hat D(\alpha)$, where $\hat D(\alpha)=\exp\left[\alpha\hat a^\dagger-\alpha^\ast \hat a\right]$ is the displacement operator.
	The photon-number statistics of this state is then recorded by a photon-number resolving detector; see, e.g., Ref. \cite{Divochiy2008}.
	This results in a sequence of $N$ measured photon numbers $\{n_{j,\alpha}\}_{j=1}^N$ for a chosen value of $\alpha$.
	By altering the complex amplitude of the coherent reference beam one gets the photon-number statistics $p_n(\alpha)$ of all coherently displaced signal states. 
	The quasiprobability $P_w(\alpha)$ is obtained locally in phase space out of the recorded data via \cite{Kiesel2012_2}
	\begin{align}\label{eq:Pwsamplingunbalanced}
		P_w(\alpha)=\sum_{n=0}^\infty \Xi_{w}(n)\,p_n(\alpha)\approx\dfrac1{N}\sum_{j=1}^N \Xi_w(n_{j,\alpha}),
	\end{align}
	where the discrete pattern function
	\begin{align}\label{eq:coeff}
		\Xi_{w}(n)&=\dfrac{2}{\pi}\int_0^{\infty}db\, b\,\Omega_w(b)\,L_n(b^2),
	\end{align}
	is related to the filter function $\Omega_w$ and $L_n(\cdot)$ are the Laguerre polynomials.
	Again, we make use of phase-independent filters.
	In close analogy with the case of balanced homodyne detection, the statistical error of Eq. \eqref{eq:Pwsamplingunbalanced} for independent-measurement events reads
	\begin{align}\label{eq:sigmaunbalanced}
		\sigma\left\{P_w(\alpha)\right\}=\dfrac1{\sqrt{N(N-1)}}\sqrt{\sum_{j=1}^N\left[\Xi_w(n_{j,\alpha})-P_w(\alpha)\right]^2}.
	\end{align}
		
\section{Experimental state regularization}
\label{ch:experiment}

	We want to notice at this point that the phase-space functions, directly sampled via Eqs. \eqref{eq:Pwsamplingbalanced} and \eqref{eq:Pwsamplingunbalanced}, follow from a numerical filtering of the recorded data after the measurement.
	This means that the sampled function $P_w$ has actually no physical existence as the $P$~function of a light field.
	Therefore, the question arises of whether light can be experimentally prepared in a state with a $P$~function equal to the filtered version $P_w$.
	This would be a realization of the filtering procedure in a quantum optical device, before any application and/or measurement of the state under consideration.
	It would also decouple the quasiprobability regularization from its determination out of measured data.
	In the following we propose and analyze an experimental scheme which solves exactly this task.
	
\subsection{Regularization of quantum states}
\label{ch:FilteredStates}	
	
	By construction the convolution kernel $\tilde\Omega_w$ in Eq. \eqref{eq:convolution}, together with Eqs. \eqref{eq:ftfilter}--\eqref{eq:smallomega}, is non-negative and normalized such that
	\begin{align}
		\int d^2\gamma\,\tilde\Omega_w(\gamma)=1.
	\end{align}
	Therefore, it corresponds to a classical probability distribution over the set of coherent amplitudes $\gamma$.
	Consequently, we can consider $P_w$ in Eq. \eqref{eq:convolution} as the $P$~function of a new state
	\begin{align}\label{eq:rhow}
		\hat\rho_w=\int d^2\alpha\,P_w(\alpha)|\alpha\rangle\langle\alpha|.
	\end{align}
	Inserting Eq. \eqref{eq:convolution} and using the Glauber-Sudarshan representation of the original state $\hat\rho$ in Eq. \eqref{eq:GS}, one obtains	
	\begin{align}\label{eq:rhofiltered}
		\hat \rho_w=\int d^2\gamma\,\tilde\Omega_w(\gamma)\,\hat D(\gamma)\hat\rho \hat D^\dagger(\gamma),
	\end{align}
	which is a mixture of the original state $\hat\rho$ undergoing various coherent displacements of amplitude $\gamma$, specified by the statistics $\tilde\Omega_w(\gamma)$.
	The underlying quantum process $\mathcal{E}_w$, which realizes the map $\hat\rho\mapsto\hat\rho_w$, is classical because all coherent input states $|\alpha\rangle$ are converted to classical output states, i.e.,
	\begin{align}
		\mathcal{E}_w(|\alpha\rangle\langle\alpha|)=\int d^2\gamma\,\tilde\Omega_w(\gamma-\alpha)|\gamma\rangle\langle\gamma|.
	\end{align}
	In the limit $w\to\infty$, the function $\tilde\Omega_w$ approaches the Dirac $\delta$ function, and, accordingly, the process leaves the input state unchanged.
	
	The problem under consideration is solved if it is possible to implement the quantum process $\mathcal{E}_w$ experimentally.
	Let us start by testing the suitability of schemes, which utilize solely linear optical components.
	It is clear, that such a process cannot be realized by a passive loss channel such as, e.g., the turbulent atmosphere \cite{Semenov2009,Vasylyev2016}, which is properly modeled as a channel with fluctuating losses.
	This becomes obvious, since a vacuum signal, $P(\alpha)=\delta(\alpha)$, is in this case always mapped to a vacuum output rather than the desired state in Eq. \eqref{eq:rhow}.
	In fact, it is necessary to include active elements---consisting of the coherent displacements---needed to realize the proper process.	
	
	It is well known that a state can be coherently displaced by a complex amplitude $\gamma$, by combining the signal beam on a highly transmissive beam splitter (field transmission $|T|\to 1$, reflection $|R|\ll 1$) with a coherent beam, to be prepared in a coherent state $|\alpha_{\mathrm{L}}\rangle$ with amplitude $\alpha_{\mathrm{L}}=T\gamma/R$.
	Obviously, the state in Eq. \eqref{eq:rhofiltered} can be produced in this manner out of the signal, by replacing the coherent reference beam by a statistical mixture of different coherent amplitudes $\alpha_{\mathrm{L}}$ following the distribution $\tilde\Omega_w$.
	This means that the reference beam must be in the classical state
	\begin{align}\label{eq:stateLO}
		\hat\rho_{\mathrm{ECF}}(w)&=\int d^2\gamma\,\tilde\Omega_w(\gamma)\,|T\gamma/R\rangle\langle T\gamma/R|,
	\end{align}
	which depends on the filter parameter $w$.
	We refer to this reference field as an engineered classical field (ECF).

	\begin{figure}[t]
		\includegraphics[clip,scale=0.9]{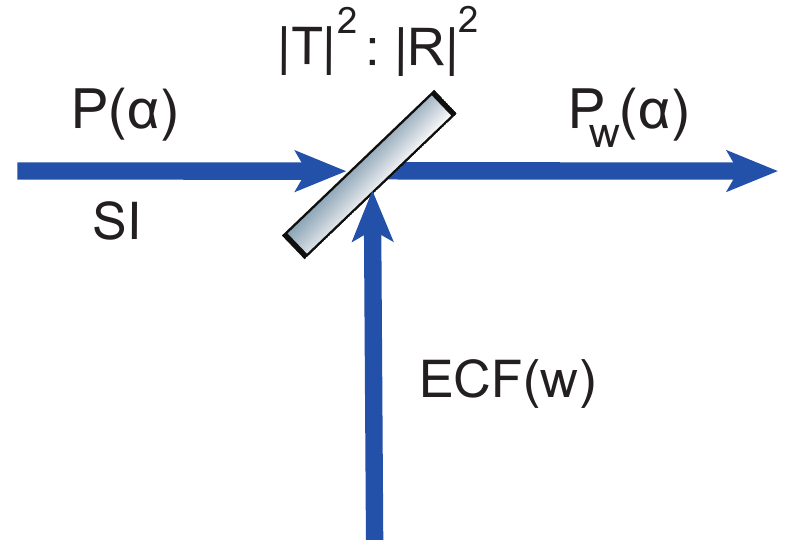}
		\caption{(Color online)
 		Linear optical implementation of the mapping procedure.
 		The $P$~function of the signal (SI) is converted to a regularized version $P_w$ by superimposing it on a highly transmissive beam splitter with a properly engineered classical field (ECF), to be adjusted by the parameter $w$. 
 	}\label{fig:setup}
	\end{figure}
	
	Our basic scenario is illustrated in Fig. \ref{fig:setup}.
	Since the transmittance of the beam splitter is close to unity, the reflection loss of the signal state can be neglected.
	In this way, any state with a singular $P$~function can be mapped to a state with a regular one.
	In particular, this allows us to provide a converter of nonclassical states to such with regular $P$~functions.
	In the limit $w\to\infty$ it holds that $\tilde\Omega_w(\gamma)\to\delta(\gamma)$, where $\delta$ is the Dirac $\delta$ function.
	Consequently, the signal state in Eq. \eqref{eq:rhofiltered} remains unaffected, i.e., $\hat\rho_w=\hat\rho$.
	The reference beam is in this case in the vacuum state; cf. Eq. \eqref{eq:stateLO}.
	Interestingly, an arbitrarily high nonfluctuating loss, which has possibly attenuated the signal and removed the negativities of $P_w$ for a chosen value of $w$, can be completely compensated by using a larger value of the filter parameter $w$.
	Such constant losses can occur, e.g., due to mode mismatching and losses in the beam splitters.
	In particular, if a loss described by the efficiency $\eta>0$ occurs, the modified filter parameter $w/\sqrt{\eta}$ must be applied to obtain a nonclassical state with a regularized $P$~function attaining negative values, in the output channel of Fig. \ref{fig:setup}.
	This can be easily shown from Eq. \eqref{eq:convolution} by applying the loss transformation, $P(\alpha)\mapsto P(\alpha/\sqrt{\eta})/\eta$, to the $P$~function.
	
	Let us mention that this phase-space regularization technique is not confined to optical systems but can be, in principle, applied to all harmonic-oscillator systems.
	A famous example are the vibrational quantum states of trapped ions \cite{Monroe1996}.
	Experimental methods for the coherent displacement, needed for the realization of our quantum process, are available for such a system; see Refs. \cite{Ziesel2013,Monroe1996}.
	
\subsection{Gaussian filtering}
\label{ch:Gauss}

	Let us start with the simplest case of Gaussian filtering [$q=2$ in Eq. \eqref{eq:Omegaq}], commonly associated with $s$-parametrized quasiprobabilities \cite{Cahill1969},
	\begin{align}\label{eq:gaussconvolution}
		P_{s}(\alpha)=\int d^2\gamma\,\tilde G_s(\gamma)\,P(\alpha-\gamma),
	\end{align}
	with a Gaussian kernel
	\begin{align}
		\tilde G_s(\gamma)=\dfrac{2}{\pi(1-s)}\exp\left[-\dfrac{2}{1-s}|\gamma|^2\right],\,\,\,\,\,\,\,\,\,\,\,\,\,\,\,\,s\leq 1.
	\end{align}
	Note that the real number $s$ plays here a similar role as the filter parameter $w$ in Eq. \eqref{eq:convolution}.
	It was shown that $P_{s}(\alpha)$ is a regular function if $s\leq 0$.
	The quantity $P_{-1}$ is also known as the Husimi $Q$ function \cite{Husimi1940}, which is non-negative and regular for all states.
	Consequently, our device in Fig. \ref{fig:setup}, in the configuration for $s\leq-1$, converts both classical and nonclassical signal states $\hat\rho$ to classical states $\hat\rho_s$ with regular and non-negative $P$~function, namely,
	\begin{align}
		 \hat\rho_s=\int d^2\alpha\,P_s(\alpha)|\alpha\rangle\langle\alpha|.
	\end{align}
	For $s=0$, Eq. \eqref{eq:gaussconvolution} yields the Wigner function \cite{Wigner1932} which is always regular, but attains negative values for a subclass of states, e.g., Fock states with the exception of the vacuum state.
	As a drawback, the nonclassicality of several states, such as squeezed vacuum states in Eq. \eqref{eq:sqstate}, cannot be uncovered on the basis of negativities of the Wigner function.
	In the case $s>0$ it can occur that $P_{s}(\alpha)$ is singular.
	Note that the Gaussian convolution in Eq. \eqref{eq:gaussconvolution} has also been used to define a measure of nonclassicality \cite{Lee1991}.
	
	As it follows from Eq. \eqref{eq:stateLO}, the ECF has to be prepared in the classical Gaussian state,
	\begin{align}
		\hat\rho_{\mathrm{ECF}}(s)=\int d^2\gamma\,\tilde G_s(\gamma)\,|T\gamma/R\rangle\langle T\gamma/R|,
	\end{align}
	for the purpose of realizing the quantum process $\hat\rho\mapsto\hat\rho_s$.
	This is basically a thermal state with the $P$~function, 
	\begin{align}
		P(\gamma)=\dfrac1{{\pi\overline{n}}}\,\exp\left(-\dfrac{|\gamma|^2}{\overline{n}}\right),
	\end{align}
	with the mean photon number
	\begin{align}
		\overline{n}=\dfrac{1-s}{2}\,\dfrac{|T|^2}{|R|^2}.
	\end{align}
	The experimental generation of a thermal state is possible, e.g., by utilizing a rotating ground glass disk; see Refs. \cite{Zavatta2007,Arecchi1965}.
	
\subsection{Non-Gaussian filtering}
\label{ch:nonGauss}

	Gaussian filtering comes along with the limitation that it does not map the $P$~function of all nonclassical states, e.g., squeezed vacuum states, to a regular quasiprobability, attaining negative values.
	This problem is completely solved by utilizing in Eq. \eqref{eq:convolution} the non-Gaussian nonclassicality filters in Eq. \eqref{eq:Omegaq} [$q>2$], discussed in Sec. \ref{ch:regularization}.
	In the limiting case of $q\to\infty$ the filter has the analytical form \eqref{eq:usedfilter}. 
	Its Fourier transform is also given by an analytical expression \cite{Kiesel2012_1},
	\begin{align}\label{eq:ftinfty}
		\tilde\Omega^{(\infty)}_w(\gamma)=\dfrac1{\pi}\dfrac{\left[J_1\left(2w|\gamma|\right)\right]^2}{|\gamma|^2},
	\end{align}
	where $J_\nu(\cdot)$ is the Bessel function of the first kind.
	For $q<\infty$, the Fourier transform can be numerically calculated as the square of the absolute value of the Fourier transform of the function in Eq. \eqref{eq:smallomega}, i.e.,
	\begin{equation}\label{eq:ftqbk}
	\begin{aligned}
		\tilde\Omega^{(q)}_w(\gamma)&=\left|\dfrac1{\pi}\int d^2\beta\,\omega_w^{(q)}(\beta)\,e^{\gamma\beta^\ast-\gamma^\ast\beta}\right|^2\\
		&=4\left(\int_0^{\infty} db\,b\,\omega^{(q)}_w(b)\,J_0(2|\gamma|b)\right)^2;
	\end{aligned}
	\end{equation}
	cf. also Fig. \ref{fig:ft}.
	It was shown in Ref. \cite{Kuehn2014} and later confirmed in experiment \cite{Agudelo2015}, that a larger value of $q$ leads to a clearly improved statistical significance of the negativity of the regularized $P$~function, $P_w$, sampled from a fixed amount of data in balanced homodyne detection.
	In fact, the significance is maximal in the limit $q\to\infty$.
	
	For our proposed experimental filtering procedure, the ECF in Fig. \ref{fig:setup} has to be prepared in the state
	\begin{align}\label{eq:stateLOq}
		\hat\rho_{\mathrm{ECF}}(w)&=\int d^2\gamma\,\tilde\Omega^{(q)}_w(\gamma)\,|T\gamma/R\rangle\langle T\gamma/R|.
	\end{align}
	In Appendix \ref{ch:techsolution} we propose a technique for the approximate preparation of the classical state in Eq. \eqref{eq:stateLOq}.
	At this point we should note that it is impossible to experimentally realize this state for $q\to\infty$ without approximations.
	This is because it requires the preparation of an unphysical state,
	\begin{align}\label{eq:stateLOinfty}
		\hat\rho_{\mathrm{ECF}}=\int d^2\gamma\,\dfrac1{\pi}\dfrac{\left[J_1\left(2w|\gamma|\right)\right]^2}{|\gamma|^2}\,|T\gamma/R\rangle\langle T\gamma/R|,
	\end{align}
	whose moments $\langle\hat a^{\dagger\ell}\hat a^k\rangle$ with $\ell+k>0$ do not exist, and, hence, e.g., this state would have an infinite energy.
	This is obvious due to the nondifferentiability of the filter $\Omega_w^{(\infty)}$ in the origin.
	However, the filters \eqref{eq:Omegaq} for $q<\infty$, corresponding to physical states in Eq. \eqref{eq:stateLOq}, approach the state for $q\to\infty$ arbitrarily close for a sufficiently large parameter $q$; compare also the curves for $q=20$ and $q=\infty$ in Fig. \ref{fig:ft}.
	
	\begin{figure}[ht]
 		\includegraphics[clip,scale=0.9]{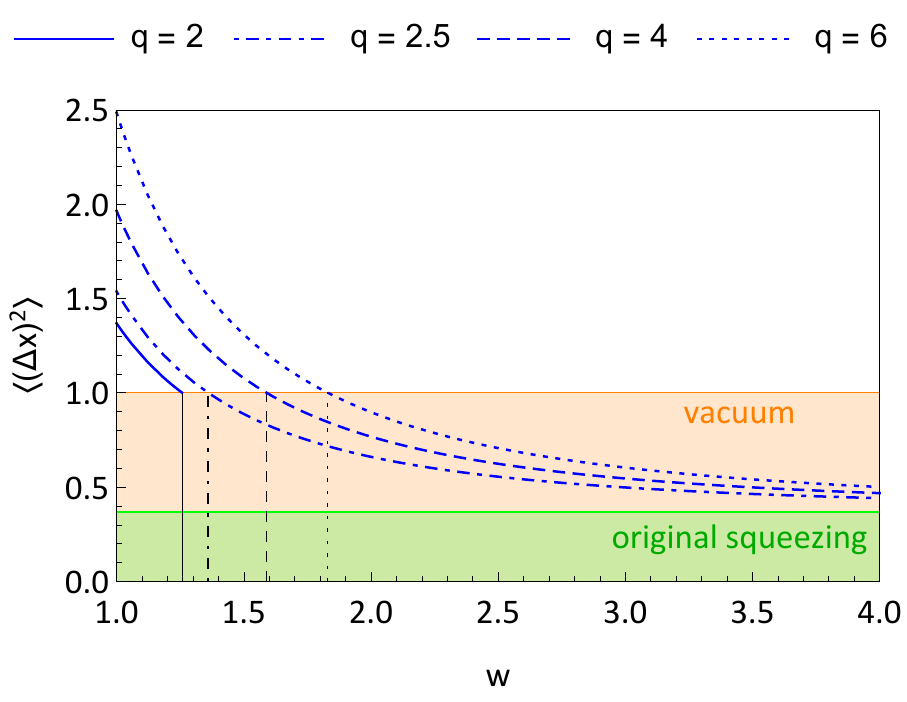}
 		\caption{(Color online)
  		The minimal quadrature variance of the generated state in Eq.~\eqref{eq:rhow} is shown as a function of the filter parameter $w$ for various values of $q$.
  		The quadrature variance of the vacuum state ($\langle[\Delta\hat x]^2\rangle_{\mathrm{vac}}=1$) and the minimal quadrature variance $\langle[\Delta\hat x]^2\rangle_{\hat \rho}\approx 0.37$ of the original squeezed vacuum state, $\hat\rho=|\xi\rangle\langle\xi|$, is also illustrated.
  		The vertical lines indicate the critical $w$ parameters $w_{\mathrm{crit}}(q)$  in Eq.~\eqref{eq:wcritbk}, corresponding to the considered $q$ parameters.
  	}\label{fig:qsq}
 	\end{figure}
	
\subsection{Application of the engineered quantum light}	
\label{ch:applications}	

	In the following we present a possible application of our technique.
	Let us consider the scenario in which the signal beam in Fig.~\ref{fig:setup} is prepared in a squeezed vacuum state $|\xi\rangle$ [cf. Eq.~\eqref{eq:sqstate}] with squeezing parameter $\xi=0.5$.
	Due to the combination with the ECF, the minimal quadrature variance of the original state is increased and hence the squeezing effect is reduced.
	Generalizing the calculation in Ref.~\cite{Saleh2012} to the broader class of nonclassicality filters studied here, the minimal quadrature variance $\langle[\Delta \hat x]^2\rangle_{\hat\rho_w}$ of the produced quantum state $\hat\rho_w$ can be explicitly determined from the minimal quadrature variance $\langle[\Delta \hat x]^2\rangle_{\hat\rho}$ of the original state $\hat\rho$ by the relation
	\begin{align}
		\langle[\Delta \hat x]^2\rangle_{\hat\rho_w}=\langle[\Delta \hat x]^2\rangle_{\hat\rho}+\dfrac{q^22^{2/q-3}}{\Gamma(2/q)}\dfrac1{w^2}.
	\end{align}
	Figure~\ref{fig:qsq} illustrates this relation as a function of the filter parameter $w$ and for different types of filter functions specified by the parameter $q$; see also Eq.~\eqref{eq:ftqbk}.
	For all values of $q$ in the range $[2,\infty)$, there exists a critical $w$-parameter
	\begin{align}\label{eq:wcritbk}
		w_{\mathrm{crit}}(q)=\dfrac{q\,2^{1/q-3/2}}{\sqrt{\Gamma(2/q)\left(1-\langle[\Delta \hat x]^2\rangle_{\hat\rho}\right)}}
	\end{align}
	such that the engineered state is quadrature squeezed for $w>w_{\mathrm{crit}}(q)$.
	Furthermore, the amount of squeezing of the original state is approached in the limit $w\to\infty$.
	For example, for $q=2.5$ and $w=4.0$ the minimal quadrature variance $0.37$ of the original state is enlarged to a variance of $0.44$, which is still well below the vacuum noise level normalized to $1$.
	Hence the resulting state shows still significant quadrature squeezing.
	
	Importantly, in the case $q>2$, corresponding to applying non-Gaussian filters, the $P$~function of the output state is regular for all finite $w$ and is thus, in principle, measurable, as demonstrated in the next section.
	Accordingly, our device converts a squeezed input state with irregular $P$~function to a quantum state which can be described by a regular $P$~function which is still squeezed.
	In contrast, the application of the Gaussian filters ($q=2$) yields either an unsqueezed output state with a regular $P$~function or a squeezed state with irregular $P$~function; see the solid blue line in Fig.~\ref{fig:qsq}.
	
	By our method, light can be prepared in nonclassical quantum states whose nonclassicality is directly visible by negativities of their regular $P$~function, without destroying the squeezing effect.
 	Squeezed light finds a manifold of applications.
 	Along with the realization of the first squeezed light sources, it was demonstrated experimentally that squeezing can improve the performance of an interferometer~\cite{Xiao1987}.
 	This ultimately leads to the application of squeezed light to highly sensitive length measurements which are necessary for the detection of gravitational waves~\cite{Grote2013}.
 	Furthermore, squeezed light was used to enhance the sensitivity of polarization measurements~\cite{Grangier1987} and intensity measurements~\cite{Xiao1988} and it was also applied in spectroscopy~\cite{Polzik1992}. 
 	For all these applications, by our method one can replace the handling of strongly irregular $P$~functions by experimentally accessible regularized ones.
		
\subsection{Multimode generalization}
\label{ch:multimode}

	The method of nonclassicality filtering was extended to multimode scenarios in Ref. \cite{Agudelo2013}.
	In particular, the $n$-mode Glauber-Sudarshan $P$~function, $P(\boldsymbol{\alpha})$ [$\boldsymbol{\alpha}=(\alpha_1,\dots,\alpha_n)^T$], of a state $\hat\rho$ is mapped to a regular function
	\begin{align}
		P_{\boldsymbol{w}}(\boldsymbol{\alpha})=\int d^{2n}\boldsymbol{\gamma}\,\tilde\Omega_{\boldsymbol{w}}(\boldsymbol{\gamma})\,P(\boldsymbol{\alpha-\gamma})
	\end{align}
	via a multimode filter function of the product form,
	\begin{align}\label{eq:multimodefilter}
		\tilde\Omega_{\boldsymbol{w}}(\boldsymbol{\gamma})=\prod_{k=1}^n\tilde\Omega_{w_k}(\gamma_k),
	\end{align}
	with adjustable width parameters $\boldsymbol{w}=(w_1,\dots,w_n)^T$.
	This corresponds to a multimode quantum process $\mathcal{E}_{\boldsymbol{w}}$ mapping the input state $\hat\rho$ to the output state
	\begin{align}
		\hat\rho_{\boldsymbol{w}}=\int d^{2n}\boldsymbol{\alpha}\,P_{\boldsymbol{w}}(\boldsymbol{\alpha})\,|\boldsymbol{\alpha}\rangle\langle\boldsymbol{\alpha}|,
	\end{align}
	with the multimode coherent states $|\boldsymbol{\alpha}\rangle=|\alpha_1\rangle\otimes\cdots\otimes|\alpha_n\rangle$.
	
	Recalling the considerations of the single-mode case, it is easy to conceive a generalization of the experimental scheme in Fig. \ref{fig:setup}.
	In fact, each mode $k=1,\dots,n$ has to be combined on a separate beam splitter with an ECF distributed according to Eq. \eqref{eq:stateLO}, depending on the individual filter parameters $w_k$.
	Note that, for using the filter in Eq. \eqref{eq:multimodefilter} all $n$ ECFs must be statistically independent, i.e., originate from different light sources.	
	
\section{Verifying the regularized states}
\label{ch:measurement}

	In this section, we describe how to determine the regularized $P$~function of the output beam in Fig. \ref{fig:setup} via balanced and unbalanced homodyne detection.
	Compared with the methods described in Sec. \ref{ch:regularization}, the unhandy filter function in the sampling formulas will now become superfluous.
	
\subsection{Balanced homodyne detection of regular states}
\label{ch:balanced}

	Recalling Eq. \eqref{eq:Pwsamplingbalanced} together with Eqs. \eqref{eq:f}--\eqref{eq:sigma}, we only have to modify the pattern function needed to sample the quasiprobability $P_w$ out of the quadrature data.
	In fact, we remove the filter function $\Omega_w$ in Eq. \eqref{eq:f} and set the upper integration limit to some positive value $b_{\mathrm{c}}=b_{\mathrm{c}}(w)$, i.e.,
	\begin{align}\label{eq:froh}
		f(\Lambda_{j,\alpha};b_{\mathrm{c}})=\dfrac{2}{\pi}\int_0^{b_{\mathrm{c}}} db\,b\,e^{b^2/2}\,\cos\left(\Lambda_{j,\alpha}b\right).
	\end{align}
	This corresponds to a rigorous postdetection filtering of the characteristic function of the $P$~function with the phase-independent rectangular function
	\begin{align}\label{eq:Omegapost}
		\Omega^{(\mathrm{post})}(\beta;b_{\mathrm{c}})=\mathrm{rect}\left(\dfrac{|\beta|}{b_{\mathrm{c}}}\right).
	\end{align}
	This procedure is well known from the filtered backprojection (see, e.g., Ref. \cite{Kak2001}), which enables the state tomography of the Wigner function via the inverse Radon transform \cite{Smithey1993}.
	It has also been applied for the first reconstruction of a state with a regular $P$~function \cite{Kiesel2008}.
	The filtering with Eq. \eqref{eq:Omegapost} is indispensable in order to suppress the sampling noise which arises due to a finite experimental data set.
	Any noise appearing in the characteristic function for $|\beta|> b_{\mathrm{c}}$ can be erased in this way, but also some state information is lost.
	In Appendix \ref{ch:systematicerror} we analyze the associated systematic errors.
	For the purpose of getting a significant result for the sampled quasiprobability $P_w$, the parameter $b_{\mathrm{c}}$ can be further increased when simultaneously increasing the number of data points.
	
	Conveniently, one can evaluate the integral in Eq. \eqref{eq:froh} to be
	\begin{equation}\label{eq:fana}
	\begin{aligned}
		&f(\Lambda_{j,\alpha};b_{\mathrm{c}})=\\
		&-\dfrac{2}{\pi}+\dfrac{2}{\pi}e^{2b_{\mathrm{c}}^2}\cos\left(2b_{\mathrm{c}}\Lambda_{j,\alpha}\right)\\
		&+\sqrt{\dfrac{2}{\pi}}e^{\Lambda_{j,\alpha}^2/2}\Lambda_{j,\alpha}\mathrm{Re}\left[\mathrm{erf}\left(\dfrac{\Lambda_{j,\alpha}+2ib_{\mathrm{c}}}{\sqrt{2}}\right)-\mathrm{erf}\left(\dfrac{\Lambda_{j,\alpha}}{\sqrt{2}}\right)\right],
	\end{aligned}
	\end{equation}
	where $\mathrm{erf}(\cdot)$ denotes the error function.
	Nicely, this pattern function is given as a closed form expression compared with Eq. \eqref{eq:f}, the latter being cumbersome to be numerically evaluated.
	Furthermore, it depends no longer on the specific shape of the employed filter function, because the filtering is completely realized before the detection.
	For further analysis we chose $b_{\mathrm{c}}=2w$ as a good compromise between systematic and statistical error.
	
	In the following, we exemplify our method for a squeezed vacuum signal field in Fig. \ref{fig:setup}, because this is the most prominent example for a nonclassical state with highly-singular $P$~function.
	To demonstrate that our technique works, we simulate for this state with squeezing parameter $\xi=0.5$ both the state manipulation, realized by the quantum process $\mathcal{E}_w$, and the measurement of the state by balanced homodyne detection. 
	The feasibility and correctness of such a simulation of quantum optics on a classical computer is comprehensively studied; see, e.g., Ref. \cite{Rahimi2016}.  
	We numerically generate $N=3\times 10^6$ random numbers for the Gaussian noise according to the classical Wigner function of the squeezed vacuum state.
		
	Following the considerations in Appendix \ref{ch:techsolution}, we prepare the ECF in Fig. \ref{fig:setup} in the classical state in Eq. \eqref{eq:LO} together with the transmission statistics in Eq. \eqref{eq:transmissioninfty}, which approximately corresponds to filtering with the function in Eq. \eqref{eq:usedfilter}. 
	The amplitude-phase modulation, which is required to realize the ECF state according to Appendix \ref{ch:techsolution}, is simulated by generating $N$ random numbers for the transmissions $\tau$ with the correct statistics.	
	The resultant quadratures obtained from the simulated balanced homodyne detection are then forwarded to the sampling formulas \eqref{eq:Pwsamplingbalanced} and \eqref{eq:sigma} together with the pattern function in Eq. \eqref{eq:fana}.
	
\begin{figure}[h]
	\centering
		\begin{minipage}[t]{1.0\linewidth}
			(a)\\
			\includegraphics[width=0.95\linewidth]{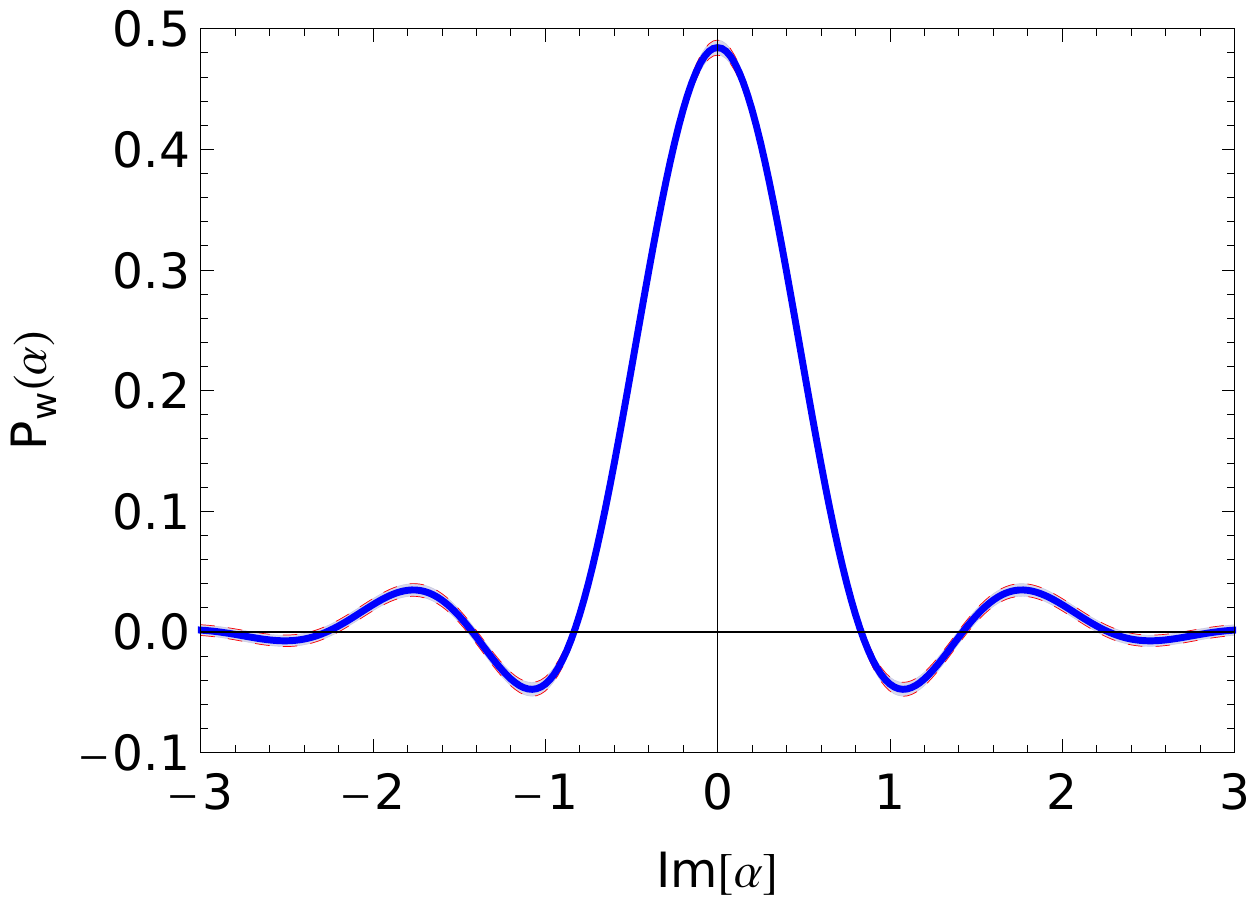}
			\\\vspace*{3ex}(b)\\
			\includegraphics[width=0.95\linewidth]{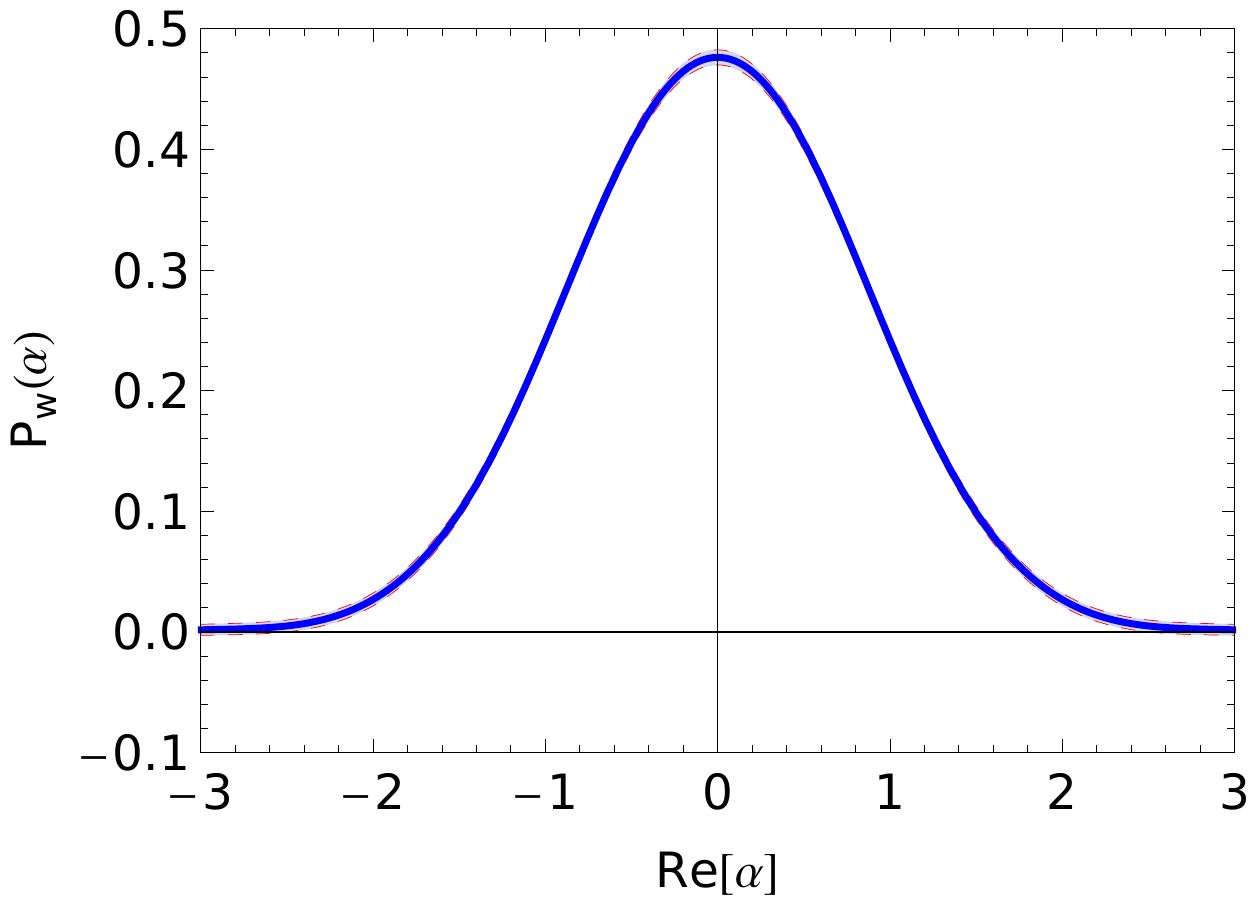}			
		\end{minipage}
		\caption{
			(Color online)
			Regularized $P$~function of a squeezed vacuum state with squeezing parameter $\xi=0.5$ for $w=1.3$, sampled from $3\times 10^6$ data points $(\phi_j,x_j)$:
			(a) squeezed axis, (b) antisqueezed axis.
			The thin dashed lines above and below the solid line correspond to an error of one standard deviation.
		}
		\label{fig:PSq}
	\end{figure}

	The sampled $P$~function for $w=1.3$ of the state under study is shown in Figs. \ref{fig:PSq} (a) and \ref{fig:PSq} (b) for the squeezed and antisqueezed axes.
	The statistical error is almost hidden behind the linewidth.
	The systematic error can be neglected because we used $\gamma_{\mathrm{c}}=100/w$ (cf. Appendix \ref{ch:systematicerror}).
	One clearly observes the typical shape of the regularized $P$~function of a squeezed vacuum state; see Ref. \cite{Agudelo2015}.
	In particular, we reach a high statistical significance of eight standard deviations for the negativity of this function.
	These negativities unambiguously certify the nonclassicality of the squeezed vacuum state.
	
\subsection{Unbalanced homodyne detection of regular states}	
\label{ch:unbalanced}
		
	A simulation of the unbalanced homodyne detection described in Sec. \ref{ch:regularization} is only possible with classical input states, whereas such a simulation for a squeezed vacuum input with nonvanishing squeezing fails; see Ref. \cite{Rahimi2016}.
	Similarly to the modification of the pattern function in balanced homodyne detection, we have to adjust the pattern function \eqref{eq:coeff}, required to obtain the regular $P$~function $P_w$ through Eq. \eqref{eq:Pwsamplingunbalanced}, as
	\begin{align}\label{eq:coeffana}
		\Xi(n;b_{\mathrm{c}})&=\dfrac{2}{\pi}\int_0^{b_{\mathrm{c}}}db\, b\,L_n(b^2).
	\end{align}
	This also corresponds to a postdetection filtering of the characteristic function of the $P$~function with the rectangular function \eqref{eq:Omegapost}.
	As in the previous section, a reasonable choice for the filter parameter $b_{\mathrm{c}}$ as a function of the parameter $w$ is $b_{\mathrm{c}}=2w$.
	Conveniently, the coefficients in Eq. \eqref{eq:coeffana} are given analytically as
	\begin{align}
		\Xi(n;b_{\mathrm{c}})&=\dfrac1{\pi}\left[L_n(b_{\mathrm{c}}^2)-L_{n+1}(b_{\mathrm{c}}^2)\right].
	\end{align}
	Last, let us mention, that it is possible to combine the displacing coherent reference beam in unbalanced homodyne detection with the ECF in Fig. \ref{fig:setup}, which realizes the filtering, into a setup with only one highly transmissive beam splitter and a displaced classical noisy reference field.
	
\section{Summary and Conclusions}
\label{ch:conclusions}
				
	Summing up, we presented a promising approach for the quantum optical implementation of a quantum process, which produces a state with a regular $P$~function out of a state with an irregular $P$~function.
	Our device uses only linear optical elements and classical auxiliary light.
	Starting with Gaussian filtering, we showed that nonclassical states with regular $P$~functions can be generated by applying non-Gaussian classical states.	
	In principle, for arbitrary quantum states, our technique allows the conversion to states with regular $P$~functions in experiment, while preserving the nonclassicality or classicality of the state under study.
	A free real parameter of the process defines the degree of filtering and allows us to generate a large set of different nonclassical states with specific properties.
	In this regard, a sufficiently high value of this parameter can even compensate high linear losses affecting the input state.
	Our technique resolves the disadvantage that the often singular $P$~function---which yields a full characterization of quantum interference effects of radiation fields---is not accessible in experiments.
	
	Beneficially, the filtering procedure happens before any detection.
	Consequently, if one aims at verifying the properties of the generated quantum state by a measurement, the nonclassicality filters do not enter in the processing of the measurement outcome.
	This drastically simplifies the data analysis, since the associated pattern functions required for balanced and unbalanced homodyne detection are given by analytical mathematical expressions.
	Moreover, the regularized state is available as a resource for further quantum technological applications.
	In this regard, we point out that the regularization of the $P$~function of a squeezed vacuum state by our non-Gaussian filters is performed without destroying the squeezing effect.
	By contrast, this is impossible by using Gaussian filtering since the regularization of the $P$~function is then achieved at the expense of the demolition of quadrature squeezing.
	We performed numerical simulations, based on the correct representation of quantum noise on a classical computer, and this proved the feasibility of our method in real experiments.
	
	Our method is not confined to continuous-variable systems, but works also in the discrete-variable regime.
	It is even easily extendible to multimode systems, which allows, e.g., the generation of entangled states with regular $P$~functions.
	Hence, we think that our technique can be useful as an important tool for the state preparation for a broad spectrum of applications.

\begin{acknowledgements}
	The authors acknowledge enlightening discussions with J. Sperling, S. Ryl, A. Luis, A. A. Semenov, and C. Di Fidio.
	This work has received funding from the European Union's Horizon 2020 research and innovation program under Grant Agreement No. 665148 (QCUMbER). 
\end{acknowledgements}

\appendix

\section{Preparation of Engineered Classical Field}
\label{ch:techsolution}

	In the following, we study a technical solution to produce an approximation of the classical state \eqref{eq:stateLOq}.
	One starts with a laser source, which emits light in a coherent state of amplitude $\gamma_{\mathrm{L}}$.
	This beam is forwarded to an amplitude and phase modulator, working as a fluctuating loss channel.
	As a result one obtains coherent states of amplitude $\gamma=\tau e^{i\phi}\gamma_{\mathrm{L}}$, with the random transmission factor $\tau\in[0,1]$ and phase shift $\phi\in[0,2\pi)$.
	In Ref. \cite{Pandey2014} such a technique was already realized in experiment via an acousto-optic modulator to generate classical light with a desired photon-number distribution out of coherent light.
	If the modulator is adjusted properly and the original laser amplitude is $\gamma_{\mathrm{L}}=T\gamma_{\mathrm{c}}/R$ with any real positive $\gamma_{\mathrm{c}}$, the desired statistical distribution $\tilde\Omega^{(q)}_w(\gamma)$ of coherent states $|T\gamma/R\rangle$ in the ECF beam in Fig. \ref{fig:setup} is mimicked for $|\gamma|\leq \gamma_{\mathrm{c}}$.
	Actually, the classical reference field is prepared in the state
	\begin{equation}\label{eq:stateLOcut}
	\begin{aligned}
		\hat\rho_{\mathrm{ECF}}&=\mathcal{N}\int_0^{\gamma_{\mathrm{c}}} dr\,r\,\tilde\Omega^{(q)}_w(r)\\
		&\times\int_0^{2\pi}d\phi\, ||T|r e^{i\phi}/|R|\rangle\langle |T|r e^{i\phi}/|R||,
	\end{aligned}
	\end{equation}
	where $\mathcal{N}$ is a suitable normalization constant and the contained pure states are coherent states.
	
	Because the distribution $\tilde\Omega^{(q)}_w(\gamma)$ only depends on $|\gamma|$, the modulator can simply use a uniform random distribution for the phase modulation, i.e., perform a full phase randomization; see also Ref. \cite{Franzen2006}.
	It remains the determination of the exact transmission distribution of the modulator.
	The latter works as an attenuator; thus, no coherent states with amplitudes greater than $|\gamma_{\mathrm{L}}|$ get out of it.
	Since the distribution $\tilde\Omega^{(q)}_w$ has infinite support, this automatically gives rise to a systematic error, which we quantify below.
	Although the reference beam is still classical in this case, and, accordingly, it cannot cause additional nonclassicality in the output in Fig. \ref{fig:setup}, it leads to an incomplete regularization of the input $P$~function. 
	However, with regard to a subsequent state reconstruction, it is for a finite amount of data not possible to distinguish a state with a regular from a state with a singular $P$~function (cf. Shannon's sampling theorem in Ref. \cite{Shannon1949}).
	That is why we speak in this situation of a partial regularization of the $P$~function.
	
\begin{figure}[h]
		\includegraphics[clip,scale=0.95]{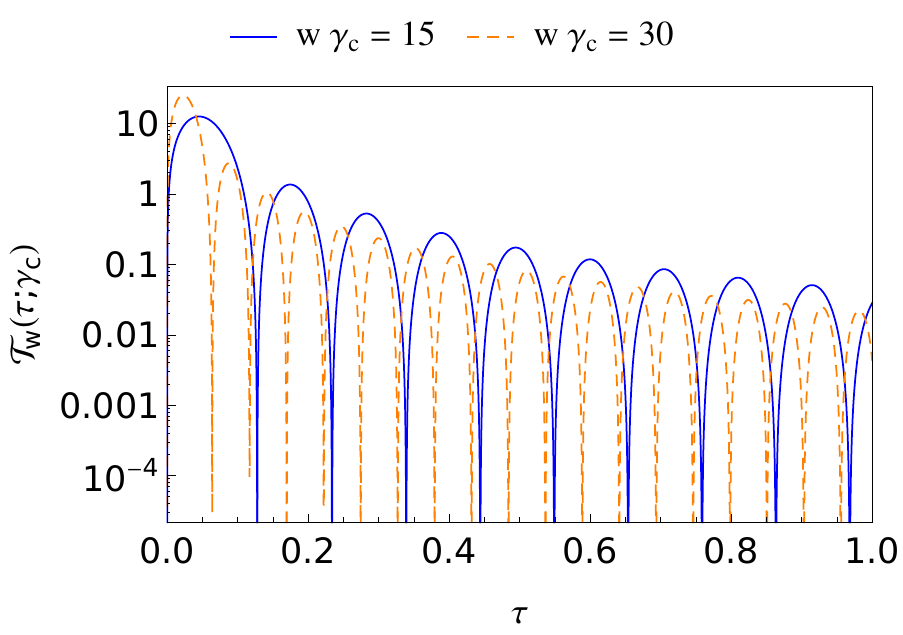}
		\caption{(Color online)
 		Transmission distribution $\mathcal{T}_w(\tau;\gamma_{\mathrm{c}})$ required for the filtering with $\Omega_w^{(\infty)}$ in Eq. \eqref{eq:usedfilter} for two different values of the product $w\gamma_{\mathrm{c}}$.
 	}\label{fig:transmissiondistribution}
	\end{figure}

	Rewriting the expression \eqref{eq:stateLOcut} as an integral over the amplitude transmission $\tau$, yields
	\begin{equation}\label{eq:LO}
	\begin{aligned}
		\hat\rho_{\mathrm{ECF}}&=\mathcal{N}\int_0^{1} d\tau\,\mathcal{T}_w(\tau;\gamma_{\mathrm{c}})\\
		&\times\int_0^{2\pi}\dfrac{d\phi}{2\pi}\, ||T|\gamma_{\mathrm{c}}\tau e^{i\phi}/|R|\rangle\langle |T|\gamma_{\mathrm{c}}\tau e^{i\phi}/|R||,
	\end{aligned}
	\end{equation}
	where 
	\begin{align}\label{eq:transmission}
		\mathcal{T}_w(\tau;\gamma_{\mathrm{c}})&= 2\pi (w\gamma_{\mathrm{c}})^2\tau\,\tilde\Omega^{(q)}_1(w\gamma_{\mathrm{c}}\tau)
	\end{align}
	is obviously the unnormalized amplitude transmission statistics to be applied to the modulator.
	It depends only on the product $w\gamma_{\mathrm{c}}$, but not explicitly on the composing variables $w$ and $\gamma_{\mathrm{c}}$.
	Remarkably, this allows us to alter the filter width parameter not only by a linear rescaling of the transmission distribution $\mathcal{T}_w$ but also by changing the amplitude $\gamma_{\mathrm{L}}$ of the laser source.
	In the limit $q\to\infty$ the transmission statistics has the analytical form
	\begin{align}\label{eq:transmissioninfty}
		\mathcal{T}_w(\tau;\gamma_{\mathrm{c}})&=\dfrac{2}{\tau}\left[J_1(2w\gamma_{\mathrm{c}}\tau)\right]^2,
	\end{align}
	which is shown in Fig. \ref{fig:transmissiondistribution} for $w\gamma_{\mathrm{c}}=15$ and $w\gamma_{\mathrm{c}}=30$.
	 
	\begin{figure}[h]
		\includegraphics[clip,scale=0.65]{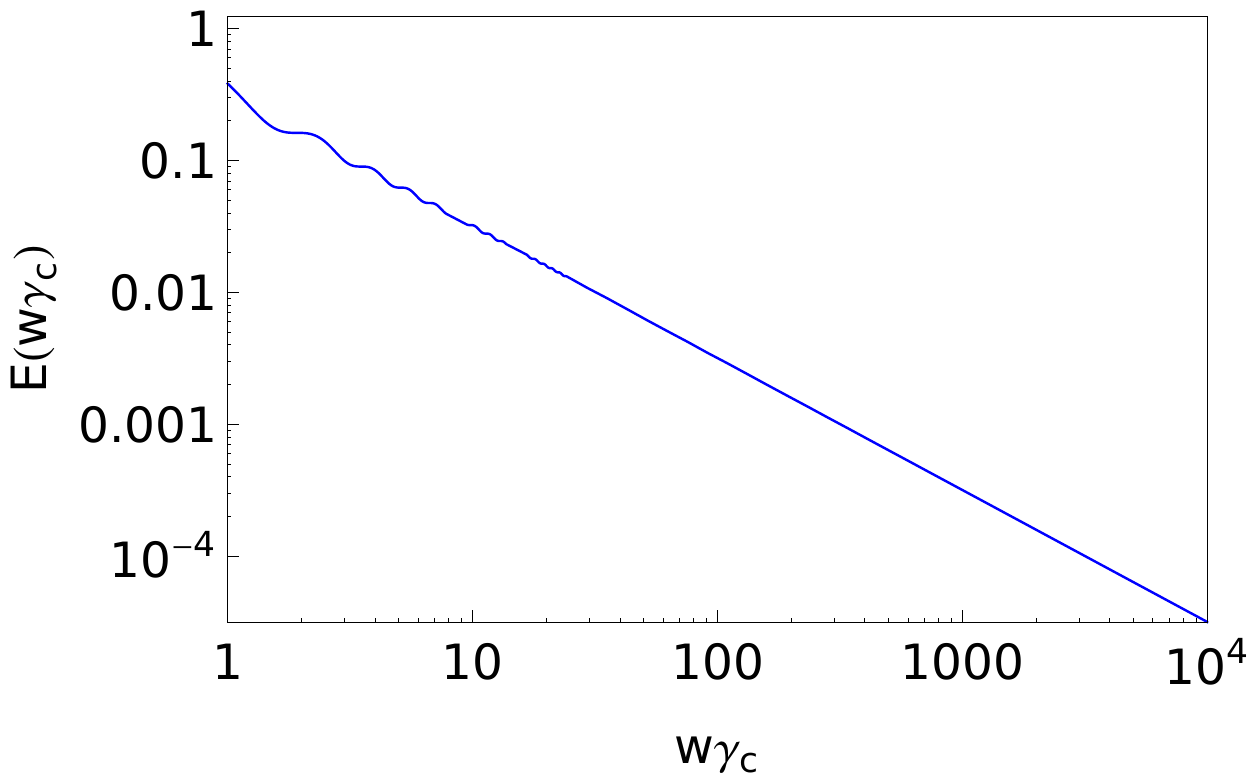}
		\caption{(Color online)
 		Systematic error $E(w\gamma_{\mathrm{c}})$ due to the finite laser amplitude $\gamma_{\mathrm{L}}$.
 	}\label{fig:error}
	\end{figure} 
	 
	Given a fixed value of the filter parameter $w$, one has to ensure a sufficiently high amplitude $\gamma_{\mathrm{L}}=T\gamma_{\mathrm{c}}/R$ of the laser source in order to minimize the systematic error
	\begin{align}\label{eq:E}
		E(w\gamma_{\mathrm{c}})&=1-\int_0^1 d\tau\,\mathcal{T}_w(\tau;\gamma_{\mathrm{c}}).
	\end{align}
	In the case of the specific transmission \eqref{eq:transmissioninfty}, we obtain
	\begin{align}\label{eq:Einfty}
		E(w\gamma_{\mathrm{c}})&=\left[J_0(2w\gamma_{\mathrm{c}})\right]^2+\left[J_1(2w\gamma_{\mathrm{c}})\right]^2.
	\end{align}
	For the parameters used in Fig. \ref{fig:transmissiondistribution}, the errors are $E(15)=2.2\%$ and $E(30)=1.1\%$.
	These imperfections, arising from a finite cut-off $\gamma_{\mathrm{c}}<\infty$, can be reduced by increasing the initial amplitude $|\gamma_{\mathrm{L}}|$ of the laser source, which results in a larger value of $\gamma_{\mathrm{c}}$.
	For instance, in the case of $\gamma_{\mathrm{c}} w=1000$ the error reduces to $E(1000)=3.2\times 10^{-4}$.
	Therefore, one should chose $w\gamma_{\mathrm{c}}$ as large as possible.
	The function $E(w\gamma_{\mathrm{c}})$ in Eq. \eqref{eq:Einfty} is illustrated in Fig. \ref{fig:error}. 
	For $w\gamma_{\mathrm{c}}\gtrapprox10$ the function $E$ can be very well approximated by $E\approx1/(\pi w\gamma_{\mathrm{c}})$.
		
	To construct a random generator, which produces the fluctuating transmission values of the modulator, it is necessary to determine the associated cumulative distribution function $F_w$ to $\mathcal{T}_w$.
	In the case of the statistics \eqref{eq:transmissioninfty}, one finds even the closed analytical expression
	\begin{align}
		F_w(\tau;\gamma_{\mathrm{c}})&=1-\left[J_0(2w\gamma_{\mathrm{c}}\tau)\right]^2-\left[J_1(2w\gamma_{\mathrm{c}}\tau)\right]^2.
	\end{align}
	Inverting this function for $\tau$, transmission values distributed according to $\mathcal{T}_w(\tau;\gamma_{\mathrm{c}})$ can be generated out of uniformly distributed random variables in $[0,1]$.

	The filter truncation, associated with the described scheme can also be considered as a filtering of the $P$~function with an effective filter
	\begin{align}\label{eq:effectivefilter}
		\tilde\Omega_w^{(q)}(\gamma;\gamma_{\mathrm{c}})=\tilde\Omega_w^{(q)}(\gamma)\cdot\mathrm{rect}(\gamma_{\mathrm{c}}-|\gamma|).
	\end{align}
	The impacts of this different filtering on the resulting filtered quasiprobability in our above proposed scheme, is described in detail in Appendix \ref{ch:systematicerror}.

\section{Systematic Errors}
\label{ch:systematicerror}

	Here we determine the systematic error arising in the quasiprobability $P_w$ with respect to the indication of nonclassicality due to the cut-off $\gamma_{\mathrm{c}}<\infty$ (cf. Appendix \ref{ch:techsolution}), and due to the postdetection filtering (cf. Sec. \ref{ch:measurement}).
	We confine ourselves to the case of filtering with $\Omega_w^{(\infty)}$ in Eq. \eqref{eq:usedfilter}.
	Including the cut-off $\gamma_{\mathrm{c}}$, the Fourier transform of the effective filter reads
	\begin{align}
		\tilde\Omega_w(\gamma;\gamma_{\mathrm{c}})=\left[\dfrac1{\sqrt{\pi}}\dfrac{J_1(2w|\gamma|)}{|\gamma|}\,\mathrm{rect}\left(\gamma_{\mathrm{c}}-|\gamma|\right)\right]^2;
	\end{align}
	cf. Eq. \eqref{eq:effectivefilter}.
	The filter itself is derived from this function by a Fourier transform, i.e.,
	\begin{equation}
	\begin{aligned}\label{eq:gammacfiltering}
		\Omega_w(\beta;\gamma_{\mathrm{c}})&=\int d^2\gamma\,\tilde\Omega_w(\gamma;\gamma_{\mathrm{c}})\,e^{\beta\gamma^\ast-\beta^\ast\gamma}\\
		&=2\int_0^{2w\gamma_{\mathrm{c}}} \dfrac{dr}{r} J_1^2(r)\,J_0\left(\dfrac{|\beta|}{w}r\right).
	\end{aligned}
	\end{equation}
	
	In both the balanced and the unbalanced homodyne detection, the sampling formula utilizes the rectangular filter in Eq. \eqref{eq:Omegapost} with $b_{\mathrm{c}}=2w$ in order to suppress the sampling noise. 
	In combination with Eq. \eqref{eq:gammacfiltering}, this leads to the total filter
	\begin{align}\label{eq:totalfilter}
		\underline{\Omega}_w(\beta;\gamma_{\mathrm{c}})&=\Omega^{(\mathrm{post})}(\beta;2w)\cdot\Omega_w(\beta;\gamma_{\mathrm{c}}).
	\end{align}
	This function has no longer a non-negative Fourier transform, which is why it may introduce negativities in the sampled quasiprobability, even if the signal state is classical.
	Therefore, we quantify this effect by the following considerations.
	
	Assume we have an arbitrary classical signal state, which has a non-negative $P$~function $P_{\mathrm{cl}}$. 
	Let us calculate the minimum of the associated regularized version
	\begin{align}
		\underline{P}_w(\alpha)&=\int d^2\gamma\,P_{\mathrm{cl}}(\gamma)\,\tilde {\underline{\Omega}}_w(\alpha-\gamma;\gamma_{\mathrm{c}}),
	\end{align}
	where $\tilde {\underline{\Omega}}_w$ denotes the Fourier transform of the filter in Eq. \eqref{eq:totalfilter} [cf. also Eq. \eqref{eq:convolution}].
	We derive for the minimum of $\underline{P}_w(\alpha)$ that
	\begin{equation}
 	\begin{aligned}
		\min_\alpha\underline{P}_w(\alpha)&=\min_\alpha\int d^2\gamma\,P_{\mathrm{cl}}(\gamma)\,\tilde {\underline{\Omega}}_w(\alpha-\gamma;\gamma_{\mathrm{c}})\\
		&\geq \int d^2\gamma\,P_{\mathrm{cl}}(\gamma)\min_\alpha \tilde{\underline{\Omega}}_w(\alpha-\gamma;\gamma_{\mathrm{c}})\\
		&=\left(\int d^2\gamma\,P_{\mathrm{cl}}(\gamma)\right)\cdot\min_\alpha \tilde{\underline{\Omega}}_w(\alpha;\gamma_{\mathrm{c}})\\
		&=\min_\alpha \tilde{\underline{\Omega}}_w(\alpha;\gamma_{\mathrm{c}}),
	\end{aligned}
	\end{equation}
	where we applied the normalization of the $P$~function in the last step.
	This result shows, that an additional fake negativity due to the filtering can be at most of the size of the minimum of the Fourier transform of the filter.
	
	\begin{figure}[h]
		\includegraphics[clip,scale=0.95]{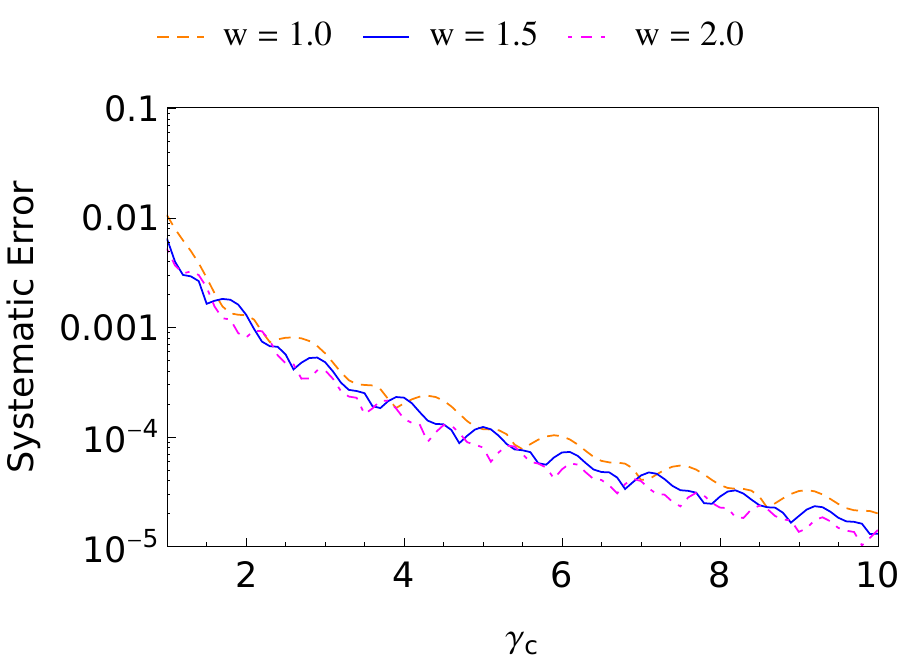}
		\caption{(Color online)
		Systematic error of the negativities of $P_w$ as a function of the cut-off amplitude $\gamma_{\mathrm{c}}$ for various values of the filter parameter $w$.
	}\label{fig:errorPOmega}
	\end{figure}	
	
	Consequently, we have to minimize $\underline{\tilde{\Omega}}_w(\gamma;\gamma_{\mathrm{c}})$ with respect to $\gamma$ in order to estimate the systematic error.
	One finds
	\begin{equation}\label{eq:fttotal}
	\begin{aligned}
		&\tilde{\underline{\Omega}}_w(\gamma;\gamma_{\mathrm{c}})\\
		&=\dfrac{4}{\pi}\int_0^{2w\gamma_{\mathrm{c}}} \dfrac{dr}{r} J_1^2(r)\int_0^{2w}db\,b\,J_0(2|\gamma|b)J_0\left(\dfrac{r}{w}b\right)\\
		&=\dfrac{8w}{\pi}\int_0^{2w\gamma_{\mathrm{c}}} \dfrac{dr}{r}\\
		&\times J_1^2(r) \dfrac{2|\gamma|J_1(4w|\gamma|)J_0(2r)-(r/w)J_0(4w|\gamma|)J_1(2r)}{4|\gamma|^2-(r/w)^2}.
	\end{aligned}
	\end{equation}
	The systematic error, as the absolute value of the minimum of \eqref{eq:fttotal}, is shown in Fig. \ref{fig:errorPOmega} as a function of $\gamma_{\mathrm{c}}$ and various values of the filter parameter $w$. 
	It decreases with increasing values of $\gamma_{\mathrm{c}}$, which can be properly controlled by the initial laser amplitude $\gamma_{\mathrm{L}}$.

\end{document}